\documentclass[conference]{IEEEtran}

\usepackage{alltt                                    
          , multirow
          , booktabs
          , listings
          , graphicx
          ,float
	,cite
          ,verbatim
         ,mathtools
	,url
	,amsmath
}
\usepackage[table]{xcolor}
\usepackage[numbers]{natbib}     
\usepackage{syntax}
\usepackage{algorithm}
\usepackage{enumitem}
\usepackage{threeparttable}
\usepackage{framed}
\usepackage{hhline}
\usepackage{algpseudocode}
\usepackage{balance}

\usepackage{expl3}
\ExplSyntaxOn
\newcommand\latinabbrev[1]{
  \peek_meaning:NTF . {
    #1\@}%
  { \peek_catcode:NTF a {
      #1., \@ }%
    {#1., \@}}}
\ExplSyntaxOff

\newcommand{\CASE}[1]{\STATE \textbf{case} #1\textbf{:} \begin{ALC@g}}
\newcommand{\ENDCASE}{\end{ALC@g}}

\newcommand{\DEFAULT}{\STATE \textbf{default:} \begin{ALC@g}}
\newcommand{\ENDDEFAULT}{\end{ALC@g}}
\newcommand{\DEFAULTLINE}[1]{\STATE \textbf{default:} }
\let\footnotesize\scriptsize

\algnewcommand{\LineComment}[1]{\State \(\triangleright\) #1}

\definecolor{LightGray}{rgb}{.9 .9 .9}

\newsavebox{\supbox}
\newcommand{\bsup}{\begin{lrbox}{\supbox}$\tt\scriptstyle}
\newcommand{\esup}{$\end{lrbox}{}^{\usebox{\supbox}}}
\def\eg{\latinabbrev{e.g}}
\def\ie{\latinabbrev{i.e}}

\definecolor{lightpurple}{rgb}{0.8,0.8,1}
\definecolor{codebg}{RGB}{255,255,255}
\definecolor{commentcolor}{RGB}{11,140,11}
\begin{document}
%

\title{RACK: Automatic API Recommendation using \\Crowdsourced  Knowledge \vspace{-.3cm}}
%
%
%
%
%

\author{\IEEEauthorblockN{Mohammad Masudur Rahman  ~~~ Chanchal K. Roy ~~~ $^\dagger$David Lo}
\IEEEauthorblockA{University of Saskatchewan, Canada, $^\dagger$Singapore Management University, Singapore\\
\{masud.rahman, chanchal.roy\}@usask.ca, $^\dagger$davidlo@smu.edu.sg}
}

\maketitle

\begin{abstract}

Traditional code search engines often do not perform well with natural language queries since they mostly apply keyword matching.
These engines thus need carefully designed queries containing information about programming APIs for code search. 
Unfortunately, existing studies suggest that preparing an effective code search query is both challenging and time consuming for the developers.
In this paper, we propose a novel API recommendation technique--RACK that recommends
a list of relevant APIs for a natural language query for code search
by exploiting keyword-API associations from the crowdsourced knowledge of Stack Overflow.
We first motivate our technique using an exploratory study with 11 core Java packages and 344K Java posts from Stack Overflow.
Experiments using 150 code search queries randomly chosen from three Java tutorial sites show that our technique recommends correct API classes within the top 10 results for about 79\% of the queries 
which is highly promising.
Comparison with two variants of the state-of-the-art technique also shows that RACK outperforms both of them not only in Top-K accuracy but also in mean average precision and mean recall by a large margin.
 
\end{abstract}


\IEEEpeerreviewmaketitle

\section{Introduction}
Studies show that software developers on average spend about 19\% of their development time in web search where they mostly look for relevant code snippets for their tasks \cite{twostudy}. 
Code search engines such as Open Hub, Koders, GitHub search and Krugle provide access to thousands of 
large open source projects which are potential sources for such snippets \cite{portfolio}.
Traditional code search engines generally employ keyword matching, \ie\ return code snippets based on lexical similarity between search query and source code. 
They expect carefully designed queries containing relevant API classes or methods from the users, and thus, often do not perform well with unstructured natural language queries.
Unfortunately, preparing an effective search query containing information about relevant APIs is not only a challenging but also a time-consuming task for the developers \cite{kevic,twostudy}.
Previous study also suggested that on average, developers with varying experience levels performed poorly in coming up with good search terms for code search \cite{kevic}.
Thus, an automated technique that translates a natural language query into a set of relevant API classes or methods (\ie\ search-engine friendly query) can greatly assist the developers in code search.
Our paper addresses this particular research problem by exploiting the crowdsourced knowledge from Stack Overflow Q \& A site.



Existing studies on API recommendation accept one or more natural language queries, and return relevant API classes and methods by analyzing
feature request history and API documentations \cite{feature}, API invocation graphs \cite{conngraph}, library usage patterns \cite{librec}, code surfing behaviour of the developers and API invocation chains \cite{portfolio}. 
\citet{portfolio} first propose \emph{Portfolio} that recommends relevant API methods for a given code search query, and demonstrates their usage from a large codebase. 
\citet{conngraph} improve upon \emph{Portfolio}
by employing further sophisticated graph-mining and textual similarity techniques.
\citet{feature} recommend relevant API methods to assist the implementation of an incoming feature request. 
Although all these techniques perform well in different working contexts, they share a set of limitations and fall short to address our research problem.
First, each of these techniques \cite{portfolio,conngraph,feature} exploits lexical similarity measure (\eg\ Dice's coefficients \cite{conngraph}) for candidate API selection. This warrants that the search query should be carefully prepared, and it should contain keywords similar to the API names. 
In other words, the developer should possess a certain level of experience on the target APIs to actually use those techniques \cite{cslog}. 
Second, API names and search queries are generally provided by different developers who may use different vocabularies to convey the same concept \cite{kevicdict}. 
Concept location community has termed it as vocabulary mismatch problem \cite{qeffect}. Lexical similarity based techniques often suffer from this problem. Hence, the performance of these techniques is not only limited but also subject to the identifier naming practices adopted in the codebase under study. 
We thus need a technique that overcomes the above limitations, and recommends relevant APIs for natural language queries from a wider vocabulary.

\begin{figure}[!t]
\centering
\includegraphics[width=3.5in]{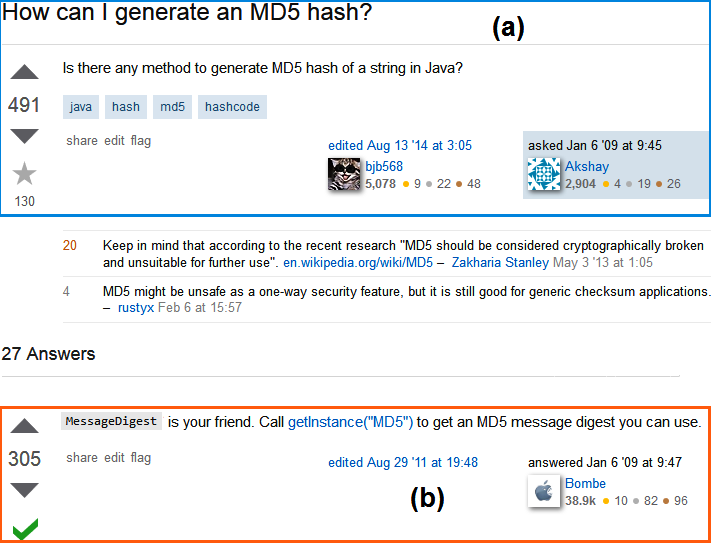}
\vspace{-.5cm}
\caption{An example of Stack Overflow (a) question \& (b) accepted answer}
\vspace{-.6cm}
\label{fig:motiv}
\end{figure}

One possible way to tackle the above challenges is to exploit crowdsourced knowledge on the usage of particular API classes and methods.
Let us consider a natural language query--\emph{``Generating MD5 hash of a Java string."} 
Now, we analyze thousands of Q \& A posts from Stack Overflow that suggest relevant APIs for this task, and then recommend APIs from them.
For instance, the Q \& A example in Fig. \ref{fig:motiv} discusses on how to generate an MD5 hash (Fig. \ref{fig:motiv}-(a)), and the accepted answer (Fig. \ref{fig:motiv}-(b)) 
suggests that \texttt{MessageDigest} API 
should be used for the task. 
Such usage of the API is also recommended by at least 305 technical users from Stack Overflow which validates the appropriateness of the usage.  
Our work is thus generic, language independent, project insensitive, and in the same time, it overcomes the vocabulary mismatch problem suffered from by the past studies.

In this paper, we propose an API recommendation technique--RACK--that exploits the association of different APIs with query keywords from Stack Overflow, and translates a natural language query for code search into a set of relevant APIs.
First, we motivate our idea of using crowdsourced knowledge for API recommendation with an exploratory study where we analyze 172,043 questions and their accepted answers from Stack Overflow. 
Second, we construct a keyword-API mapping database using those questions and answers where the keywords (\ie\ programming requirements) are extracted from questions and the APIs (\ie\ programming solutions) are collected from corresponding accepted answers.
Third, we propose an API recommendation technique that employs two heuristics on keyword-API associations, and recommends a ranked list of API classes for a given query.   
The baseline idea is to capture and learn the responses from millions of technical users (\eg\ developers, researchers, programming hobbyists) for different programming problems, and then exploit them for relevant API recommendation. 
Our technique (1) does not rely on the lexical similarity between query and source code of projects for API selection, and (2) addresses the vocabulary mismatch problem by using a significantly large vocabulary (\ie\ 20K) produced by millions of users of Stack Overflow. 
Thus, it has a great potential to overcome the challenges faced with the past studies.     
   
An exploratory study with 344,086 Java related posts from Stack Overflow shows that (1) each post uses at least two different API classes on average, and (2) about 65\% of the classes from each of the 11 core Java API packages are used in those posts.
This suggests the potential of Stack Overflow for relevant API recommendation.
Experiments using 150 code search queries randomly chosen from three Java tutorial sites show that our technique can recommend relevant APIs with a Top-10 accuracy of about 79\% which is highly promising.
We also compared with two variants of the state-of-the-art technique by \citet{feature}, and report that our technique outperformed both of them not only in Top-K accuracy but also in mean average precision and mean recall by a large margin.
Thus, the paper makes the following contributions:
\begin{itemize}
\item An exploratory study that suggests the potential of Stack Overflow for relevant API recommendation for code search using natural language queries.
\item A keyword-API mapping database that maps 655K question keywords to 551K API classes from Stack Overflow.
\item A novel technique that exploits query keyword-API associations from crowdsourced knowledge, and translates a natural language query into a set of relevant API classes.
\item Comprehensive evaluation of the proposed technique with four metrics, and comparison with the state-of-the-art.
\end{itemize}

\section{Exploratory Study}\label{sec:exploratory}
Our technique relies on the mapping between natural language keywords from the questions of Stack Overflow and API classes from corresponding accepted answers for translating a code search query into relevant API classes. 
Thus, an investigation is warranted if such answers contain any API related information and the questions contain any query keywords.
We perform an exploratory study using 344K posts from Stack Overflow, and analyze the usage and coverage of standard Java API classes in those posts. 
We also explore if the question titles are a potential source for keywords for code search.
We particularly answer three research questions as follows:    
\begin{itemize}
\item \textbf{RQ$\mathbf{_1}$}: To what extent do the accepted answers from Stack Overflow refer to standard Java API classes?
\item \textbf{RQ$\mathbf{_2}$}: To what extent  are the API classes from each of the core Java packages covered (\ie\ mentioned) in the accepted answers from Stack Overflow?
\item \textbf{RQ$\mathbf{_3}$}: Do the titles from Stack Overflow questions contain potential keywords for code search?
\end{itemize}

\begin{table}
\centering
\caption{API Packages for Exploratory Study}\label{table:packages}
\vspace{-.2cm}
\resizebox{2.5in}{!}{%
\begin{threeparttable}
\begin{tabular}{l|c||l|c}
\hline
\textbf{Package} & \textbf{\#Class} & \textbf{Package} & \textbf{\#Class}\\
\hline
\texttt{java.lang} & 255 & \texttt{java.net} & 84\\
\hline
\texttt{java.util} & 470 & \texttt{java.security} & 148\\
\hline
\texttt{java.io} & 105 & \texttt{java.awt} & 423\\
\hline
\texttt{java.math} & 09 & \texttt{java.sql} & 29\\
\hline
\texttt{java.nio} & 189 & \texttt{java.swing} & 1,195\\
\hline
\texttt{java.applet} & 05 & &\\
\hline
\end{tabular}
\centering
\end{threeparttable}
}
\vspace{-.6cm}
\end{table}

\subsection{Data Collection}\label{sec:datacoll}
We collect 172,043 questions and their accepted answers from Stack Overflow using StackExchange data explorer \cite{de} for our investigation. Since we are interested in Java APIs, we only collect such questions that are tagged as \emph{java}. 
Besides, we apply several other constraints--(1) each of the questions should have at least three answers (\ie\ average answer count) with one answer being accepted as solution, in order to ensure that the questions are answered substantially and successfully, and (2)
the accepted answers should contain code like elements such as code snippets or code tokens so that API information can be extracted from them. 
We identify the code elements with the help of \texttt{<code>} tags in the HTML source of the answers (details in Section \ref{sec:apiextract}), and use Jsoup \cite{jsoup}, a popular Java library, for HTML parsing and content extraction.

We collect a total of 2,912 Java classes from 11 core API packages\footnote{https://en.wikipedia.org/wiki/Java_package} of standard Java edition 6, one of the most stable versions \cite{stable}, for our study. The goal is to find out if these classes are referred to in Stack Overflow posts, and if yes, to what extent they are referred to.
We first use Reflections \cite{reflection}, a runtime meta data analysis library, to collect the API classes from JRE 6, and then apply regular expressions on their fully qualified names for extracting the class name tokens.
Table \ref{table:packages} shows class statistics of the 11 API packages selected for our investigation.

We also collect a set of 18,662 real life search queries (of the first author) from Google for over the last eight years which are analyzed to answer the third research question. 
Although the queries come from a single user, 
they contain a large vocabulary of 9,029 distinct natural language keywords, and the vocabulary is built over a long period of time.
Thus, a study using those queries can produce significant intuitions.

\subsection{API Class Name Extraction} \label{sec:apiextract}
Several existing studies \cite{peter,robi,email} extract code elements such as API packages, classes and methods from unstructured natural language texts (\eg\ forum posts, mailing lists) using information retrieval (\eg\ TF-IDF) and island parsing techniques.
In the case of island parsing, they apply a set of regular expressions describing Java language specifications \cite{gosling}, and isolate the land (\ie\ code elements) from water (\ie\ free-form texts). 
We borrow their parsing technique \cite{peter}, and apply it to the extraction of API elements from Stack Overflow posts. 
Since we are interested in the API classes referred to in the posts, we adopt a selective approach for identifying them.
We first isolate the code like sections from HTML source of each of the answers from Stack Overflow using \texttt{<code>} tags. Then we split the sections based on white spaces and punctuation marks, and collect the tokens having the camel-case notation for Java class (\eg\ \texttt{HashSet}). 
According to the existing studies \cite{peter,robi}, such parsing of code elements sometimes might introduce false positives.
Thus, we restrict our exploratory analysis to a closed set of 2,912 API classes from 11 core Java packages (Table \ref{table:packages}) for avoiding false positives (\eg\ camel-case tokens but not valid API classes). 

\begin{figure}[!t]
\centering
\includegraphics[width=3.5in]{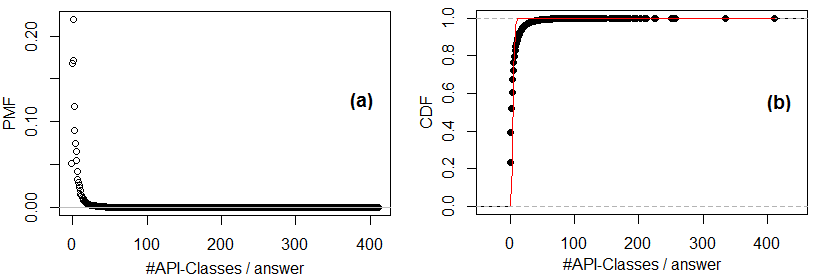}
\vspace{-.6cm}
\caption{API frequency distribution-- (a) API frequency PMF (b) API frequency CDF}
\vspace{-.6cm}
\label{fig:api-pmf-cdf}
\end{figure}

\subsection{Answering RQ$_1$: API use in accepted SO answers}
Since our API recommendation technique exploits keyword-API associations from Stack Overflow, we investigate if the accepted answers actually use certain APIs of our interest in the first place.
According to our investigation, out of 172,043 accepted answers, 136,796 (79.51\%) answers refer to one or more Java classes (\ie\ standard API or user defined), and 104,983 (61.02\%) of them use API classes from the 11 core Java packages (Table \ref{table:packages}) as a part of their solution.

\begin{figure}[!t]
\centering
\includegraphics[width=3.5in]{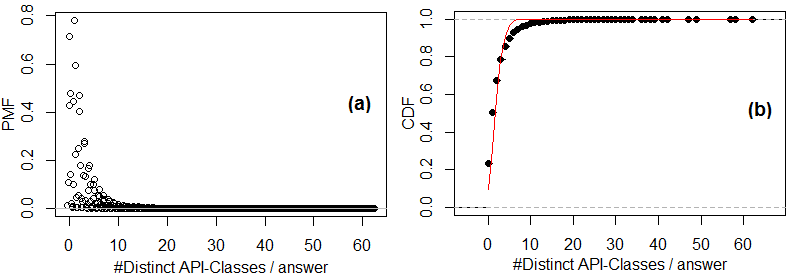}
\vspace{-.6cm}
\caption{Distinct API frequency distribution-- (a) Distinct API frequency PMF (b) Distinct API frequency CDF}
\vspace{-.4cm}
\label{fig:dapi-pmf-cdf}
\end{figure}

Fig. \ref{fig:api-pmf-cdf} shows (a) probability mass function (PMF) and (b) cumulative density function (CDF) for the total frequency of API classes used in each of the Stack Overflow answers where the classes belong to the core Java packages.
Both density curves suggest that the frequency observations derive from a heavy-tailed distribution, and majority of the densities accumulate over a short frequency range.
The empirical CDF curve also closely matches with the theoretical CDF \cite{cdf} (\ie\ red curve in Fig. \ref{fig:api-pmf-cdf}-(b)) of a Poisson distribution. 
Thus, we believe that the observations are probably taken from a Poisson distribution.
We get a 95\% confidence interval over [5.08, 5.58] for mean frequency ($\lambda=5.32$) which suggests that the API classes from the core packages are referred to at least five times on average in each of the answers from Stack Overflow.
We also get 10$^{th}$ quantile at frequency=2  and 97.5$^{th}$ quantile at frequency=10  which suggest that only 10\% of the frequencies are below 3 and only 2.5\% of the frequencies are above 10.
Fig. \ref{fig:dapi-pmf-cdf} shows similar density curves for the frequency of distinct API classes in each of the accepted answers.
We get a 95\% confidence interval over [2.33, 2.41] for mean frequency ($\lambda=2.37$) which suggests that at least two distinct classes are used on average in each answer.
30$^{th}$ quantile at frequency = 1 and 80$^{th}$ quantile at frequency = 4 suggest that 30\% of the Stack Overflow answers refer to at least one API class whereas 20\% of the answers refer to at least four distinct API classes from the core Java packages. 

Thus, to answer \textbf{RQ$\mathbf{_1}$}, at least two API classes from the core packages are referred to in each of the accepted answers from Stack Overflow that contain API classes from those packages, and they are used at least five times on average in each answer.

\begin{figure}[!t]
\centering
\includegraphics[width=2.6in]{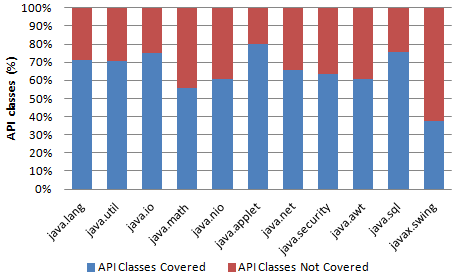}
\vspace{-.3cm}
\caption{Coverage of API classes from core packages by Stack Overflow answers}
\vspace{-.6cm}
\label{fig:api-class-coverage}
\end{figure}

\begin{figure}[!t]
\centering
\includegraphics[width=2.6in]{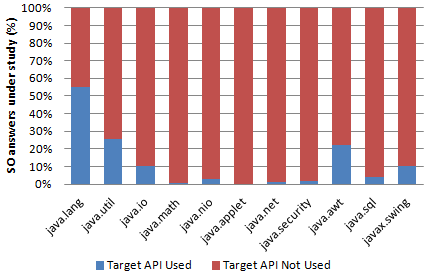}
\vspace{-.3cm}
\caption{Use of core packages in Stack Overflow answers}
\vspace{-.7cm}
\label{fig:api-coverage-so}
\end{figure}

\subsection{Answering RQ$_2$: API coverage in accepted answers}\label{sec:apicov}
Since our technique exploits mapping between API classes in Stack Overflow answers and keywords from corresponding questions for API recommendation, we need to investigate if such answers actually use 
a significant portion of the API classes from the core packages. 
We thus identify the occurrence of the classes from core packages in Stack Overflow answers, and determine API coverage for those packages.

Fig. \ref{fig:api-class-coverage} shows the fraction of the classes that are used in Stack Overflow answers for each of the 11 core packages under study.
We note that at least 60\% of the classes are used in Stack Overflow for nine out of 11 packages.
The remaining two packages--\texttt{java.math} and \texttt{javax.swing} have 55.56\% and 37.41\% class coverage respectively.
Among those nine packages, three large packages-- \texttt{java.lang}, \texttt{java.util} and \texttt{java.io} even have a class coverage over 70\%.
Fig. \ref{fig:api-coverage-so} shows the fraction of Stack Overflow answers (under study) that use API classes from each of the core 11 packages.  
We note that classes from \texttt{java.lang} package are used in over 50\% of the answers, which is quite expected since the package contains the frequently used and basic classes such as \texttt{String, Integer, Method, Exception} and so on.
Two packages-- \texttt{java.util} and \texttt{java.awt} that focus on utility functions (\eg\ unzip, pattern matching) and user interface controls (\eg\ radio button, check box) respectively  have a post coverage over 20\%. 
We also note that classes from \texttt{java.io} and \texttt{javax.swing} packages are used in over 10\% of the Stack Overflow answers, whereas such statistic for the remaining six packages is less than 10\%.

Thus, to answer \textbf{RQ$\mathbf{_2}$}, on average, about 65.15\% of the API classes from each of the core Java packages are used in Stack Overflow answers, and at least 12.22\% of the answers refer to the classes from each single API package as a part of their solutions. 
These findings suggest a high potential of Stack Overflow for API recommendation.

\subsection{Answering RQ$_3$: Search keywords in SO questions}
Our technique relies on the mapping between natural language tokens from Stack Overflow questions and API classes from corresponding accepted answers for translating a code search query into several relevant API names. 
Thus, we need to investigate if the texts from such questions actually contain keywords used for code search or not. 
We are particularly interested in the title of a Stack Overflow question since it summarizes the technical requirement of the question using a few words, and also quite resembles a search query.
We analyze the titles of 172,043 Stack Overflow questions and 18,662 real life queries used for Google search.
Since we are interested in code search queries, we only select those queries that contain any of these keywords--\emph{java, code, example} and \emph{programmatically} for our analysis.
A search using such keywords in the query is generally intended for code example search. We get 1,703 such queries containing 1,461 distinct natural language tokens from our query collection. 

\begin{figure}[!t]
\centering
\includegraphics[width=2.1in]{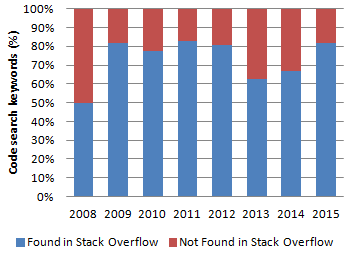}
\vspace{-.3cm}
\caption{Coverage of keywords from the collected queries  in Stack Overflow questions}
\vspace{-.4cm}
\label{fig:qtoken-so}
\end{figure}

\begin{figure}[!t]
\centering
\includegraphics[width=3.5in]{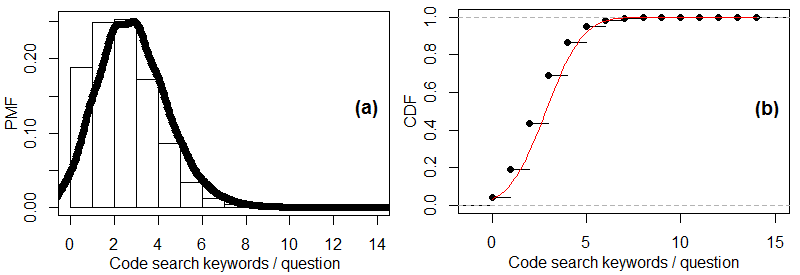}
\vspace{-.6cm}
\caption{Collected search query keywords in Stack Overflow-- (a) Keyword frequency PMF (b) Keyword frequency CDF}
\vspace{-.6cm}
\label{fig:qtoken-pmf-cdf}
\end{figure}

According to our analysis, the question titles contain 20,391 unique tokens after performing natural language processing (\ie\ stop word removal, splitting and stemming), and the tokens match 66.94\% of the keywords collected from our code search queries.
Fig. \ref{fig:qtoken-so} shows the fraction of the search keywords that match with the tokens from Stack Overflow questions for the past eight years starting from 2008. 
We note that on average, 73.03\% of the code search keywords from each year match with Stack Overflow tokens.
Such statistic reaches up to 80\% for the year 2009 to year 2011. One possible explanation for this is that the user (\ie\ first author) was a professional developer then, and most of the queries were programming or code example related.
Fig. \ref{fig:qtoken-pmf-cdf} shows  (a) probability mass function, and (b) cumulative density function for keyword frequency in the question titles.
We note that the density curve shows central tendency like a normal curve (\ie\ bell shaped curve), and the empirical CDF also closely matches with the theoretical CDF (\ie\ red curve) of a normal distribution with $\mu=2.85$ and $\sigma=1.54 $.
Thus, we believe that the frequency observations come from a normal distribution. We get a mean frequency, $\mu=2.85$ with 95\% confidence interval over [2.84, 2.86], which suggests that each of the question titles from Stack Overflow contains approximately three code search keywords on average.

Thus, to answer \textbf{RQ$\mathbf{_3}$}, titles from Stack Overflow questions contain a significant amount of the keywords that were used for real life code search.
Each title contains approximately three query keywords on average, and their tokens match with about 73\% of our collected code search keywords when considered on a yearly basis. 

\begin{figure*}[!t]
\centering
\includegraphics[width=7in]{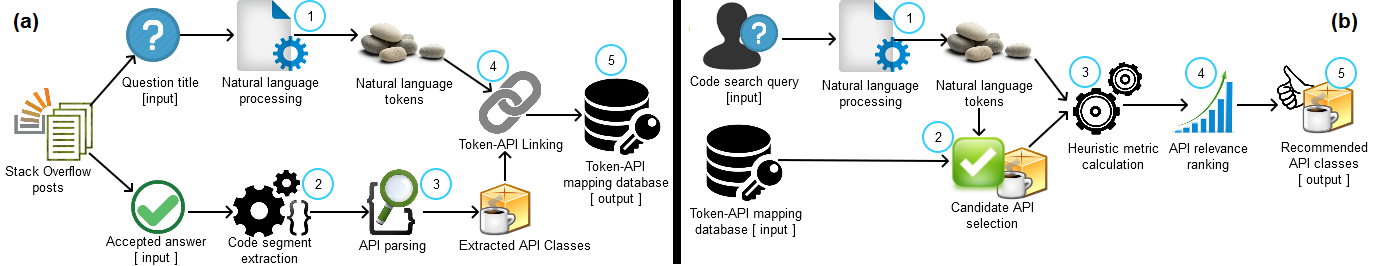}
\vspace{-.2cm}
\caption{Proposed technique for API recommendation--(a) Construction of token-API mapping database, (b) Translation of a code search query into relevant API classes}
\vspace{-.6cm}
\label{fig:sysdiag}
\end{figure*}

\section{RACK: Automatic API Recommendation using Crowdsourced  Knowledge}\label{sec:proposed}
According to the exploratory study (Section \ref{sec:exploratory}), at least two API classes are used in each of the accepted answers of Stack Overflow, and about 65\% of  the API classes from the core packages are used in those answers.
Besides, the titles from Stack Overflow questions are a major source for code search keywords. Such findings suggest that Stack Overflow might be a potential source for code search keywords and API classes relevant to them.
Since we are interested in exploiting this keyword-API association from Stack Overflow questions and answers for API recommendation, we need a technique that stores such associations, mines them automatically, and then recommends the most relevant APIs. 
Thus, our proposed technique has two major steps -- (a) Construction of token-API mapping database, and (b) Recommendation of relevant APIs for a code search query. 
Fig. \ref{fig:sysdiag} shows the schematic diagram of our proposed technique--RACK--  for API recommendation.

\subsection{Construction of Token-API Mapping Database}\label{sec:dbdevelop}
Since our technique relies on keyword-API associations from Stack Overflow, we need to extract and store such associations for quick access.
In Stack Overflow, each question describes a technical requirement such as \emph{``how to send an email in Java?''} The corresponding answer offers a solution containing  code example(s) that refer(s) to one or more API classes (\eg\ \texttt{MimeMessage}, \texttt{Transport}).
We capture such requirement and API classes carefully, and exploit their semantic association for the development of token-API mapping database.
Since the title summarizes a question using a few words, we only use the titles from the questions.
Besides, acceptance of an answer by the person who posted the question indicates that the answer actually meets the requirement in the question.
Thus, we consider only the accepted answers from the answer collection for our analysis. 
The construction of the mapping database has several steps as follows:

\textbf{Token Extraction from Titles:} We collect title(s) from each of the questions, and apply standard natural language pre-processing steps such as stop word removal, splitting and stemming on them (Step 1, Fig. \ref{fig:sysdiag}-(a)).
Stop words are the frequently used words (\eg\ the, and, some) that carry very little semantic for a sentence.
We use a stop word list \cite{stop} hosted by Google for the stop word removal step.
The splitting step splits each word containing any punctuation mark (\eg\ ?,!,;), and transforms it into a list of words. 
Finally, the stemming step extracts the root of each of the words (\eg\ ``send" from ``sending") from the list, where Snowball stemmer \cite{prompter,codes} is used.
Thus, we extract a set of unique and stemmed words that collectively convey the semantic of the question title, and we consider them as the tokens from the title.  

\textbf{API Class Extraction:} We collect the accepted answer for each of the questions, and parse their HTML content using Jsoup parser \cite{jsoup} for code segments (Step 2, 3, Fig. \ref{fig:sysdiag}-(a)).
We extract all \texttt{<code>} tags from the content as they generally contain code segments \cite{surfclipse}. 
It should be noted that code segments may sometimes be demarcated by other tags or no tag at all.
However, identification of such code segments is challenging and often prone to false-positives. Thus, we restrict our analysis to contents inside \texttt{<code>} tags for code segment collection from Stack Overflow.
We split each of the segments based on punctuation marks and white spaces, and discard the programming keywords. Existing studies \cite{email,peter} apply island parsing for API method or class extraction where they use a set of regular expressions.
Similarly, we use a regular expression for Java class \cite{gosling}, and extract the API class tokens having camel case notation.  
Thus, we collect a set of unique API classes from each of the accepted answers. 

\textbf{Token-API Linking:} Natural language tokens from a question title hint about the technical requirement described in the question, and API names from the accepted answer represent the relevant APIs that can meet such requirement.  
Thus, the programming Q \& A site--Stack Overflow-- inherently provides an important semantic association between a list of tokens and a list of APIs. 
For instance, our technique generates a list of natural language tokens--\{\emph{generat, md5, hash}\}-- and an API token-- \texttt{MessageDigest}-- from the showcase example on MD5 hash (Fig. \ref{fig:motiv}).
We capture such associations from 136,796 Stack Overflow question and accepted answer pairs, and store them in a relational database (Step 4, 5, Fig. \ref{fig:sysdiag}-(a)) for relevant API recommendation for any code search query.

\subsection{API Relevance Ranking \& Recommendation}\label{sec:relranking}
In the token-API mapping database, each token associates with different APIs, and each API associates with a number of tokens.
Thus, we need a technique that carefully analyzes such associations, identifies the candidate APIs, and then recommends the most relevant ones from them for a given query.
It should be noted that we do not apply the traditional association rule mining since our investigations suggest that many token and API sets extracted from our constructed database (Section \ref{sec:dbdevelop}) have low support.
Thus, the mined rules might not be sufficient for API recommendation. The API ranking and recommendation involve several steps as follows:

\subsubsection{Identification of Keyword Context}\label{sec:adjacent}
In natural language processing, the context of a word refers to the list of other words that co-occur with that word in the same phrase, sentence or even the same paragraph \cite{cooccur}.
Co-occurring words complement the semantics of one another \cite{rada}. 
\citet{wordsim} analyze programming posts and tags from Stack Overflow Q \& A site, and use word context for determining semantic similarity between any two software-specific words.
In this research, we identify words that co-occur with each keyword in the thousands of question titles from Stack Overflow. 
For each keyword, we refer to these co-occurring words as its \emph{context}. We then opportunistically use these context words for estimating semantic relevance between any two keywords.

\subsubsection{Candidate API Selection} In order to collect candidate APIs for a query,  we employ two heuristics. These heuristics consider not only the association between keywords and APIs but also the coherence among the keywords.
Thus, the key idea is to identify such APIs as candidates that are both likely for the query keywords and functionally consistent to one another.

\textbf{Keyword-API Co-occurrence (KAC):} Stack Overflow discusses thousands of programming problems, and the discussions contain various natural language keywords and reference to a number of APIs.
According to our observation, several keywords might co-occur with a particular API or a particular keyword might co-occur with several APIs across different programming solutions.
This co-occurrence generally takes place either by chance or due to semantic relevance.
Thus, if carefully analyzed, such co-occurrences could be a potential source for semantic association between keywords and APIs.
We capture these co-occurrences (\ie\ associations) between keywords and APIs, discard the random associations using a heuristic threshold ($\delta$), and then collect the top APIs ($L[K_i]$) for each keyword ($K_i$) that co-occurred most frequently with the keyword at Stack Overflow.  
\begin{equation*}
\setlength{\abovedisplayskip}{.3em}
\setlength{\belowdisplayskip}{.3em}
L[K_i]=\{A_j\mid A_j\epsilon A ~\wedge~ rank_{freq}(K_i\rightarrow A_j)\le \delta\} 
\end{equation*}
Here, $K_i\rightarrow A_j$ denotes the association between a keyword $K_i$ and an API $A_j$, 
$rank_{freq}$ returns rank of the association from the ranked list based on association frequency, and $\delta$ is a heuristic rank threshold.
In our research, we consider top five (\ie\ $\delta=5$) APIs for each keyword which is chosen based on iterative experiments on our dataset.



\textbf{Keyword-Keyword Coherence (KKC):}  
Since a code search query might contain multiple keywords (\ie\ describing a single task), the candidate APIs should be not only relevant to multiple keywords but also consistent with one another.
\citet{wordsim} determine semantic similarity between any two software specific words by using their contexts from Stack Overflow questions and answers.
We adapt their technique for identifying coherent keyword pairs which are then used for collecting candidate APIs functionally relevant to those pairs.
We (1) develop a context ($C_i$) for each of the $n$ query keywords by collecting its co-occurred words from thousands of question titles from Stack Overflow, (2) determine semantic similarity for each of the $^nC_2$ keyword pairs based on their context derived from Stack Overflow, and (3) use those measures to identify the coherent pairs and then to collect the functionally relevant APIs for them.
At the end of this step, we have a set of APIs for each pair of coherent keywords.

Suppose, two query keywords $K_i$ and $K_j$ have context word list $C_i$ and $C_j$ respectively. 
Now, the candidate APIs ($L_{coh}$) that are relevant to both keywords and functionally consistent with one another can be selected as follows:
\begin{equation*}
\setlength{\abovedisplayskip}{.3em}
\setlength{\belowdisplayskip}{.3em}
L_{coh}[K_i,K_j]=\{L[K_i]\cap L[K_j] \mid cos(C_i,C_j)>\gamma\}
\end{equation*}
Here, $cos(C_i,C_j)$ denotes the \emph{cosine similarity} \cite{surfclipse} between two context lists, and $\gamma$ is the similarity threshold. 
We consider $\gamma=0$ in this work based on iterative experiments on our dataset.
$L[K_i]$ and $L[K_j]$ are top frequent APIs for the two keywords-- $K_i$ and $K_j$.
Thus, $L[K_i,K_j]$ contains such APIs that are relevant to both keywords (\ie\ co-occurred with them at Stack Overflow answers), and functionally consistent with one another given that the target keywords are coherent themselves (\ie\ semantically similar).
\vspace*{-.2cm}
\begin{algorithm}
\caption{API Relevance Ranking Algorithm}
\label{algo}
\begin{algorithmic}[1]
\Procedure{RACK}{$Q$}\Comment{$Q$: code search query}
\State $R \gets$ \{\}\Comment{list of relevant APIs}
\LineComment{collecting keywords from the search query}
\State $K \gets$ collectKeywords($Q$)
\LineComment{collecting candidate APIs}
\State $L \gets$ getKACList($K$)
\State $L_{coh} \gets$ getKKCList($K$)
\LineComment{estimating relevance of the candidate APIs}
\For{Keyword $K_i$ $\in$ $K$}
\State $sortedAPIs \gets$ sortByFreq($L[K_i]$)
\For{APIClass $A_j \in sortedAPIs$}
\LineComment{likelihood score of an API}
\State $S_{KAC} \gets$ getKACScore($A_j,sortedAPIs$) 
\State $R[A_j].score \gets R[A_j].score + S_{KAC}$
\EndFor
\EndFor
\For{Keyword $K_i, K_j$ $\in$ $K$}
\State $C_i \gets$ getContextList($K_i$)
\State $C_j \gets$ getContextList($K_j$)
\LineComment{relevance of an API with multiple keywords}
\State $S_{KKC} \gets$ getKKCScore($C_i,C_j$)
\For{APIClass $A_j \in L_{coh}[K_i,K_j]$}
\State $R[A_j].score \gets R[A_j].score + S_{KKC}$
\EndFor
\EndFor
\LineComment{ranking of the APIs}
\State $rankedAPIs \gets$ sortByScore($R$)
\State \textbf{return} $rankedAPIs$ 
\EndProcedure
\end{algorithmic}
\end{algorithm}
\setlength{\textfloatsep}{2pt}

\begin{table*}
\centering
\caption{An Example of API Recommendation using RACK}\label{table:example}
\vspace{-.2cm}
\resizebox{5.8in}{!}{%
\begin{threeparttable}
\begin{tabular}{l|c|c|l|c|c|l|c|c|l|c}
\hline
\multirow{2}{*}{\textbf{java}} & \multicolumn{2}{c|}{\textbf{Scores}} & \multirow{2}{*}{\textbf{parser}} & \multicolumn{2}{c|}{\textbf{Scores}} & \multirow{2}{*}{\textbf{html}} & \multicolumn{2}{c|}{\textbf{Scores}} & \textbf{Recommended} & \textbf{Total}\\
\hhline{~--~--~--~~}
 & S$_{KAC}$& S$_{KKC}$ &  & S$_{KAC}$& S$_{KKC}$ &  & S$_{KAC}$& S$_{KKC}$ & \textbf{APIs}  & \textbf{Score} \\
\hline
\texttt{List} & 1.00 & 0.20 & \texttt{Document}& 1.00& 0.42 &\texttt{Document}  &1.00 & & \texttt{Document} & 2.42 \\
\hline
\texttt{ArrayList} & 0.80 & & \texttt{List}& 0.80& & \texttt{Jsoup} &0.80 & & \texttt{File} & 2.10\\
\hline
\texttt{File} & 0.60 & 0.20 & \texttt{Element} & 0.60&0.42 & \texttt{Element} & 0.60& & \texttt{List} & 2.00\\
\hline
\texttt{Map} & 0.40 & & \texttt{File} & 0.40& 0.42 & \texttt{Elements} & 0.40&  & \texttt{Element} &1.62 \\
\hline
\texttt{Runnable} & 0.20 & & \texttt{Node} & 0.20& & \texttt{File} &0.20 & 0.28  &\texttt{Jsoup} &0.80 \\
\hline
\end{tabular}
\centering
\end{threeparttable}
}
\vspace{-.6cm}
\end{table*}

\vspace*{-.3cm}
\subsubsection{API Relevance Ranking Algorithm} Fig. \ref{fig:sysdiag}-(b) shows the schematic diagram, and Algorithm \ref{algo} shows the pseudo code of our API relevance ranking algorithm.
Once a search query is submitted, we (1) perform Parts-of-Speech (POS) tagging on the query for extracting the meaningful terms such as nouns and verbs \cite{autocomment}, and (2) apply standard natural language processing (\ie\ stop word removal, splitting, and stemming) on them to extract the stemmed words (Line 4, Algorithm \ref{algo}).
For example, the query--\emph{``html parser in Java"} turns into three keywords--\emph{`html', `parser'} and \emph{`java'} at the end of the above step.
We then apply the two heuristics--\emph{KAC} and \emph{KKC}-- on those stemmed keywords, and collect candidate APIs from the token-API linking database (Step 2, Line 5--Line 7).
The candidate APIs are selected based on not only their co-occurrence with the query keywords but also the coherence among the keywords.  
We then use the following two metrics (derived from the above heuristics) to estimate the relevance of the candidate APIs for the query.

\textbf{API Likelihood} estimates the probability of co-occurrence of a candidate API ($A_j$) with an associated keyword ($K_i$) from the search query. It considers the rank of the API in the ranked list based on keyword-API co-occurrence frequency (\ie\ $KAC$), and then provides a normalized score as follows.
\begin{equation*}
\setlength{\abovedisplayskip}{.3em}
\setlength{\belowdisplayskip}{.3em}
S_{KAC}(A_j, K_i)=1-\frac{rank(A_j, sortByFreq(L[K_i]))}{|L[K_i]|}
\end{equation*}
Here, $S_{KAC}$ denotes the API Likelihood estimate, and it ranges from 0 (\ie\ not likely at all for the keyword) to 1 (\ie\ very much likely for the keyword).

\textbf{API Coherence} estimates the relevance of a candidate API ($A_j$) to multiple keywords from the query simultaneously.
Since the search query targets a particular programming task, each of the recommended APIs should be relevant to multiple keywords from the query.
One way to heuristically determine such relevance is to exploit the semantic similarity between the corresponding keywords that co-occurred with that API.
We thus determine semantic similarity between any two keywords ($K_i, K_j$) using their context lists ($C_i,C_j$) \cite{wordsim}, and then propagate that measure to each of the candidate APIs ($A_j$) that
co-occurred with both keywords (\ie\ $KKC$) as a proxy to relevance between the candidate and the two keywords.
\begin{equation*}
\setlength{\abovedisplayskip}{.3em}
\setlength{\belowdisplayskip}{.3em}
S_{KKC}(A_j,K_i,K_j)=cos(C_i,C_j) \mid (K_i\rightarrow A_j)\wedge (K_j\rightarrow A_j)
\end{equation*}
Here, $S_{KKC}$ denotes the API Coherence estimate, and it ranges from 0 (\ie\ not relevant at all with multiple keywords) to 1 (\ie\ very much relevant).
It should be noted that each candidate, $A_j$, comes from $L[K_i]$ or $L[K_j]$, \ie\ the API is already relevant (\ie\ frequently co-occurred) to each of $K_i$ and $K_j$ in their corresponding contexts. $S_{KKC}$ investigates how similar those contexts are, and thus heuristically estimates the relevance of the API, $A_j$, to both keywords.

We first compute \emph{API Likelihood} for each of the candidate APIs that suggests the likeliness of the API for each keyword from the query (Line 9--Line 16).  
Then we determine \emph{API Coherence} for each candidate API that suggests relevance of the API to multiple keywords from the query (Line 17--Line 25).
Once both metrics are calculated for each of the entries from $L$ and $L_{coh}$ (Step 3, Fig. \ref{fig:sysdiag}-(b)), the scores are accumulated for each individual candidate API (Line 14 and Line 23, Algorithm \ref{algo}).
The candidates are then ranked based on their accumulated scores, and top K APIs from the list are returned for recommendation (Line 26--Line 28, Algorithm \ref{algo}, Step 4, 5, Fig. \ref{fig:sysdiag}-(b)).

\textbf{Example:} Table \ref{table:example} shows a working example on how our API recommendation technique--RACK-- works.
We first collect the top 5 (\ie\ $\delta$) candidate APIs for each of the keywords--\emph{`java',`parser'} and \emph{`html'}-- based on their co-occurrence frequencies with the keywords.
Then we calculate the likelihood (\ie\ S$_{KAC}$) of each candidate. For example, \texttt{Document} has a maximum likelihood of 1.00 among the candidates for both \emph{'parser'} and \emph{'html'}.
We then collect coherence (\ie\ S$_{KKC}$) of each candidate API based on semantic relevance among the keywords. For example, \emph{`parser'} and \emph{`html'} have a semantic relevance of 0.42 on the scale from 0 to 1, and that score is added to 
the overlapping candidates--\texttt{Document}, \texttt{Element} and \texttt{File}-- between these two keywords. We then accumulate both scores for each candidate, 
discard the duplicate candidate APIs (\ie\ superclass or subclass),
and finally get a ranked list.
From the list, we see that \texttt{Document}, \texttt{Element} and \texttt{Jsoup} are highly relevant APIs for the query--\emph{``java parser html"}.

\section{Experiment}\label{sec:experiment}
One of the most effective ways to evaluate a technique for API recommendation is to analyze the relevance of the recommended APIs for a target query.
We evaluate our technique using 150 code search queries, determine its performance using four metrics, and 
then compare with two variants of the state-of-the-art technique.
We particularly answer four research questions through our experiments as follows:
\begin{itemize}
\item \textbf{RQ$\mathbf{_4}$:} How does the proposed technique perform in recommending relevant APIs for a code search query?
\item \textbf{RQ$\mathbf{_5}$:} How effective are the proposed heuristics--KAC and KKC--in capturing the relevant APIs for a query? 
\item \textbf{RQ$\mathbf{_6}$:} Is our selection of keywords from a given query effective in retrieving the relevant APIs? 
\item \textbf{RQ$\mathbf{_7}$:} Can RACK outperform the state-of-the-art technique in recommending APIs for any given set of queries?
\end{itemize}

\subsection{Experimental Dataset}\label{sec:dataset}
\textbf{Data Collection:} We collect 150 code search queries for our experiment from three Java tutorial sites-- KodeJava \cite{kodejava}, JavaDB \cite{javadb} and Java2s \cite{java2s}.
These sites discuss hundreds of programming tasks that involve the usage of different APIs from the standard Java libraries.
Each of these task descriptions generally has three parts--(1) a title (\ie\ question) for the task, (2) one or more code snippets, and (3) an associated prose explaining the code. 
The title (\eg\ {``How do I decompress a GZip file in Java?"}) summarizes the programming task in natural language using a few keywords, and it quite resembles a query for code search as well.
We thus use such titles as the code search queries for our experiment in this research.

\textbf{Gold Set Development:} The prose explaining the code often refers to one or more APIs (\eg\ \texttt{GZipOutputStream}, \texttt{FileOutputStream}) from the code snippet(s) that are found essential for the task. 
In other words, such APIs can be considered as the most relevant ones for the target task.
We collect such APIs from the prose against each of the task titles from our dataset, and develop a gold set--\emph{API-goldset}--for the experiment.
Since relevance of the APIs is determined based on working code examples and their associated prose from the publicly available popular tutorial sites, 
the subjectivity associated with the relevance is minimized \cite{conngraph}.

\subsection{Performance Metrics}\label{sec:metrics}
We choose four performance metrics for evaluation and validation that are widely used by relevant literature \cite{portfolio,conngraph,feature}.
Two of them are related to recommendation systems whereas the rest two come from information retrieval domain.

\textbf{Top-K Accuracy}: It refers to the percentage of the search queries for which at least one API is correctly recommended within the Top-K results by a recommendation technique. Top-K Accuracy can be defined as follows:
\begin{equation*}
\setlength{\abovedisplayskip}{0em}
\setlength{\belowdisplayskip}{0em}
Top\textendash K Accuracy (Q) = \frac{\sum_{q\in Q} isCorrect(q, Top\textendash K)}{|Q|} \time 100\%
\end{equation*}
Here, $isCorrect(q, Top\textendash K)$ returns a value 1 if there exists at least one API from the \emph{API-gold set} in the Top-K results, and returns 0 otherwise.
$Q$ denotes the set of all search queries.

\textbf{Mean Reciprocal Rank@K (MRR@K)}: Reciprocal rank@K refers to the multiplicative inverse of the rank of the first relevant API in the Top-K results returned by a technique.
Mean Reciprocal Rank@K (MRR@K) averages such measures for all search queries in the dataset.

\textbf{Mean Average Precision@K (MAP@K)}: \emph{Precision@K} calculates the precision at the occurrence of every single relevant API in the ranked list. \emph{Average Precision@K (AP@K)} averages the \emph{precision@K} for all relevant APIs in the list for a code search query. \emph{Mean Average Precision@K} is the mean of \emph{average precision@K} for all queries from the dataset.


\textbf{Mean Recall@K (MR@K)}: Recall@K refers to the percentage of gold set APIs that are correctly recommended for a code search query in the Top-K results by a technique. Mean Recall@K (MR@K) averages such measures for all queries.

\subsection{Evaluation of RACK}\label{sec:results}
Each of the queries summarizes a programming task that demands the use of one or more APIs from standard Java libraries.
Our technique recommends the top 10 relevant APIs for each query which are then compared with the \emph{API-goldset} for evaluation and validation using the above four metrics. 

\begin{table}[!t]
\centering
\caption{Experimental Results}\label{table:result}
\vspace{-.2cm}
\resizebox{3.3in}{!}{%
\begin{threeparttable}
\begin{tabular}{l|c|c|c}
\hline
\textbf{Performance Metric} & \textbf{Top-3} & \textbf{Top-5} & \textbf{Top-10} \\
\hline
Top-K Accuracy & 49.33\% & 62.67\% & \textbf{78.67}\%  \\
\hline
Mean Reciprocal Rank@K (MRR@K) & 0.17 & 0.17 & 0.17  \\
\hline
Mean Average Precision@K & 30.39\% & 33.36\% & \textbf{34.92}\%  \\
\hline
Mean Recall@K (MR@K) & 23.71\% & 33.48\% & \textbf{45.02}\% \\
\hline
\end{tabular}
\centering
\end{threeparttable}
}
\vspace{-.3cm}
\end{table}

\begin{figure}[!t]
\centering
\includegraphics[width=2.5in]{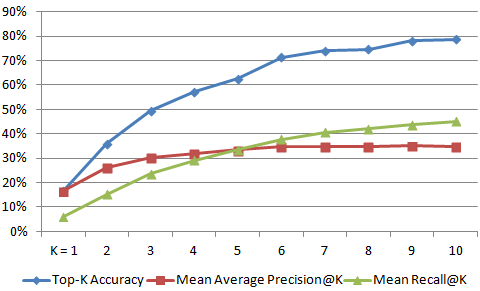}
\vspace{-.3cm}
\caption{Top-K Accuracy, Mean Average Precision@K, and Mean Recall@K}
\vspace{-.5cm}
\label{fig:perfk110}
\end{figure}

\begin{table}[!t]
\centering
\caption{Role of Different Heuristics}\label{table:component}
\vspace{-.2cm}
\resizebox{3.2in}{!}{%
\begin{threeparttable}
\begin{tabular}{l|l|c|c|c}
\hline
\textbf{Heuristics} & \textbf{Metric} & \textbf{Top-3} & \textbf{Top-5} & \textbf{Top-10}\\
\hline
& Accuracy & 50.00\% & 66.00\% & 78.00\%\\
\hhline{~----}
\{Keyword-API & MRR@K & 0.18 & 0.18 & 0.18 \\
\hhline{~----}
 Co-occurrence (KAC)\}& MAP@K & 31.44\% & 34.99\% & \textbf{35.41}\%  \\
\hhline{~----}
& MR@K & 23.99\% & 34.20\% & 44.80\%  \\
\hline
\hline
& Accuracy & 34.00\% & 39.33\% & 39.33\% \\
\hhline{~----}
\{Keyword-Keyword & MRR@K & 0.15 & 0.15 & 0.15\\
\hhline{~----}
 Coherence (KKC)\} & MAP@K & 22.78\% & 24.08\% & 24.11\% \\
\hhline{~----}
& MR@K & 15.24\% & 19.02\% & 19.52\% \\
\hline
\hline
\{Keyword-API & Accuracy &  49.33\% & 62.67\% & \textbf{78.67}\%  \\
\hhline{~----}
 Co-occurrence (KAC) \& & MRR@K & 0.17 & 0.17 & 0.17  \\
\hhline{~----}
Keyword-Keyword & MAP@K & 30.39\% & 33.36\% & \textbf{34.92}\%  \\
\hhline{~----}
 Coherence (KKC)\} & MR@K & 23.71\% & 33.48\% & \textbf{45.02}\% \\
\hline
\end{tabular}
\centering
\end{threeparttable}
}
\end{table}

Table \ref{table:result} shows the performance details of our technique for Top-3, Top-5 and Top-10 API recommendation.
We see that the technique recommends correctly for about 79\% of the queries with a mean average precision of 34.92\% and a mean recall of 45.02\% which are highly promising, especially the Top-K accuracy and the recall, according to relevant literature \cite{conngraph,portfolio}.
While the technique provides a recommendation accuracy of 63.00\% for Top-5 results, precision and recall remain close to 33.50\%.
However, for Top-10 recommendation, the accuracy and recall measures increase significantly, and the precision remains comparable. 
We do not notice any change in mean reciprocal rank (MRR) for different Top-K recommendations by our technique.
Fig. \ref{fig:perfk110} shows how different performance metrics--accuracy, precision and recall change over different values of $K$.
We also see that each of these metrics becomes stationary at $K=10$, which actually supports our choice of $K$ values for top result collection.

Thus, to answer \textbf{RQ$\mathbf{_4}$}, our API recommendation technique--RACK recommends correct APIs for about 79\% of the queries with a precision of 34.92\% and a recall of 45.02\% on average.

We investigate effectiveness of the two applied heuristics-- KAC and KKC, and justify their combination in the API ranking algorithm (Algorithm \ref{algo}).
Table \ref{table:component} demonstrates how effective each of the heuristics is in capturing relevant APIs for a given code search query. 
We see that our technique recommends correctly for 78.00\% of the queries with 35.41\% precision and 44.88\% recall when KAC is considered in isolation.
On the other hand, the technique provides at most 40.00\% accuracy for Top-10 recommendation with KKC heuristic considered in isolation.
However, our technique performs the best when both heuristics are used in combination. 
It provides a maximum of about 79.00\% recommendation accuracy with 34.92\% precision and 45.02\% recall for Top-10 results.

Thus, to answer \textbf{RQ$\mathbf{_5}$}, KAC is found more effective than KKC in capturing relevant APIs. However, combination of both heuristics provides the maximum performance.
Thus, their combination for API ranking might be justified.

Since our technique identifies relevant APIs based on their co-occurrence with query keywords, the keywords from each query should be chosen carefully.
We extract noun and verb terms from the query (Section \ref{sec:relranking}), and use them for our experiment.
In this section, we investigate if the selection of such terms from the query is effective or not.   
Table \ref{table:bias} summarizes our comparative analyses using different set of queries.
We see that our technique does not perform well especially for Top-3 and Top-5 results when all terms from a search query are used.
The performance improves when only \emph{noun} terms are considered from the query. 
However, the accuracy and the recall for Top-10 results do not reach the maximum.
The performance is also not much interesting when only \emph{verb} terms are considered.
However, our technique performs the best especially in terms of accuracy and recall when both the \emph{noun} and the \emph{verb} terms are used together.


Thus, to answer \textbf{RQ$\mathbf{_6}$}, important keywords from a query mainly consist of its noun and verb terms, and our query term selection is found quite effective in retrieving relevant APIs.

\begin{table}[!t]
\centering
\caption{Effect of Different Query Term Selection}\label{table:bias}
\vspace{-.2cm}
\resizebox{3in}{!}{%
\begin{threeparttable}
\begin{tabular}{l|l|c|c|c}
\hline
\textbf{Query Terms} & \textbf{Metric} & \textbf{Top-3} & \textbf{Top-5} & \textbf{Top-10}\\
\hline
\multirow{4}{*}{} & Accuracy & 48.00\% & 57.33\% & 78.00\% \\
\hhline{~----}
All terms & MRR@K & 0.17 & 0.17 & 0.17\\
\hhline{~----}
from query& MAP@K & 29.67\% & 31.40\% & 33.67\% \\
\hhline{~----}
& MR@K & 22.71\% & 30.29\% & 43.67\% \\
\hline
\hline
\multirow{4}{*}{Noun terms only} & Accuracy & 50.00\% & 65.33\% & 72.67\% \\
\hhline{~----}
 & MRR@K & 0.22 & 0.22 & 0.22\\
\hhline{~----}
& MAP@K & 33.17\% & 36.56\% & \textbf{36.79}\% \\
\hhline{~----}
& MR@K & 24.71\% & 34.33\% & 41.60\% \\
\hline
\hline
\multirow{4}{*}{Verb terms only} & Accuracy & 18.67\% & 23.33\% & 26.67\% \\
\hhline{~----}
 & MRR@K & 0.07 & 0.07 & 0.07\\
\hhline{~----}
 & MAP@K & 11.44\% & 12.71\% & 13.23\% \\
\hhline{~----}
& MR@K & 7.94\% & 11.11\% & 12.61\% \\
\hline
\hline
\multirow{4}{*}{} & Accuracy  &  49.33\% & 62.67\% & \textbf{78.67}\%  \\
\hhline{~----}
 Noun and Verb & MRR@K &  0.17 & 0.17 & 0.17  \\
\hhline{~----}
{terms combined} & MAP@K & 30.39\% & 33.36\% & \textbf{34.92}\%  \\
\hhline{~----}
& MR@K & 23.71\% & 33.48\% & \textbf{45.02}\% \\
\hline
\end{tabular}
\centering
\end{threeparttable}
}
\vspace{-.4cm}
\end{table}

\begin{table}[!t]
\centering
\caption{Comparison with Existing Techniques}\label{table:compare}
\vspace{-.2cm}
\resizebox{3in}{!}{%
\begin{threeparttable}
\begin{tabular}{l|l|c|c|c}
\hline
\textbf{Technique} & \textbf{Metric} & \textbf{Top-3} & \textbf{Top-5} & \textbf{Top-10}\\
\hline
\multirow{4}{*}{\citet{feature}-I}& Accuracy & 30.00\% & 38.67\% & 42.00\% \\
\hhline{~----}
 & MRR@K & 0.19 & 0.19 & 0.19\\
\hhline{~----}
& MAP@K & 23.33\% & 24.62\% & 23.53\% \\
\hhline{~----}
& MR@K & 13.50\% & 18.94\% & 25.89\% \\
\hline
\hline
\multirow{4}{*}{\citet{feature}-II} & Accuracy & 30.67\% & 37.33\% & \textbf{48.67}\% \\
\hhline{~----}
 & MRR@K & 0.17 & 0.17 & 0.17\\
\hhline{~----}
& MAP@K & 23.00\% & 23.77\% & 23.47\% \\
\hhline{~----}
& MR@K & 14.78\% & 21.06\% & \textbf{33.44}\% \\
\hline
\hline
\multirow{4}{*}{} & Accuracy  & 49.33\% & 62.67\% & \textbf{78.67}\%  \\
\hhline{~----}
 RACK & MRR@K &   0.17 & 0.17 & 0.17  \\
\hhline{~----}
{(Proposed technique)} & MAP@K & 30.39\% & 33.36\% & \textbf{34.92}\%  \\
\hhline{~----}
& MR@K & 23.71\% & 33.48\% & \textbf{45.02}\% \\
\hline
\end{tabular}
\centering
\end{threeparttable}
}
\end{table}

\begin{figure}[!t]
\centering
\includegraphics[width=2.5in]{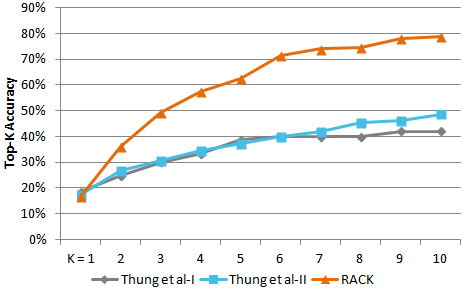}
\vspace{-.3cm}
\caption{Top-K Accuracy comparison with existing techniques}
\label{fig:compare-topk}
\end{figure}

\subsection{Comparison with Existing Techniques}\label{sec:compare}
\citet{feature} take in a feature request and return a list of relevant API methods both by mining of feature request history and by analysis of textual similarity between the request and the API documentations of those methods.
To the best of our knowledge, this is the latest closest study to our work, and thus, we select it for comparison.
Since feature request history is not available in our experimental settings, we implement \emph{Description-Based Recommender} module from the technique.
We collect API documentations of 3,300 classes from the Java standard libraries (\ie\ JDK 6), and develop Vector Space Model (VSM) for each of the classes.
In fact, we develop two models for each API class using (1) class header comments only, and (2) both class header and method header comments, and implement two variants-- \citeauthor{feature}-I and \citeauthor{feature}-II for our experiment.
We use \emph{Apache Lucene} \cite{lucene} for VSM development and for textual similarity matching between the API classes and each of the queries from our dataset.

\begin{figure}[!t]
\centering
\includegraphics[width=2.6in]{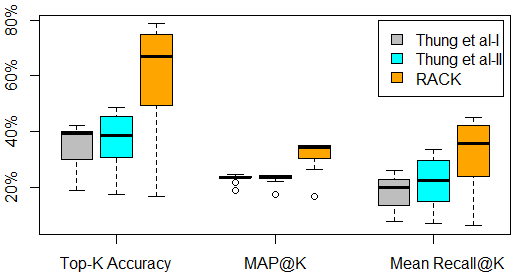}
\vspace{-.2cm}
\caption{Comparison with existing techniques}
\vspace{-.1cm}
\label{fig:compare}
\end{figure}

Table \ref{table:compare} summarizes the comparative analysis between our technique--RACK-- and the two variants of \citeauthor{feature}.
We see that the variants can provide a maximum of about 49\% accuracy with 23.47\% precision and 33.44\% recall for Top-10 results. 
On the other hand, RACK provides a maximum accuracy of 79\% with 34.92\% precision and 45.02\% recall which are significantly higher.
We investigate how the Top-K accuracy changes over different K values for each of these three techniques.
From Fig. \ref{fig:compare-topk}, we see that accuracy for RACK increases gradually up to 79\% whereas such performance measures for the textual similarity based techniques stop at 49\%.
From Fig. \ref{fig:compare}, we see that RACK performs significantly better than both variants in terms of all three metrics-- accuracy, precision and recall. 
Our median accuracy is above 70\% whereas such measure for those variants is below 40\%. The same goes for precision and recall measures.
Thus, all the findings above suggest that textual similarity between query and API signature or documentations might not be always effective for API recommendation.
Our technique overcomes that issue through applying two co-occurrence based heuristics--KAC and KKC-- which analyzes the crowdsourced knowledge from Stack Overflow.
Performance reported for \citeauthor{feature} is project-specific, and the technique is restricted to feature requests \cite{feature}. 
On the contrary, our technique is generic and adaptable for any type of code search. It is also independent of any subject systems. 
More importantly, it exploits the expertise of a large crowd of technical users for API recommendation which was not considered by the past studies.
Thus, our technique possibly has a greater potential.


Thus, in order to answer \textbf{RQ$\mathbf{_7}$}, our proposed technique-- RACK-- outperforms two variants of the state-of-the-art technique for API recommendation in terms of Top-K accuracy, precision and recall by a large margin.

\section{Threats to Validity}\label{sec:threats}\vspace{-.1cm}
Threats to internal validity relate to experimental errors and biases \cite{wordsim}.
We develop \emph{API gold set} for each query by analyzing the code examples and the discussions from tutorial sites which might involve some subjectivity.
However, each of the examples is a working solution to the corresponding task (\ie\ query), and they are frequently consulted.
Thus, the gold set development from sample code is probably a more objective evaluation approach than human judgements of API relevance that introduces more subjective bias \cite{conngraph}. 
According to the exploratory findings (Section \ref{sec:apicov}), our technique
might be effective only for the recommendation of popular and frequently used APIs. Since fully qualified names are mostly missing in Stack Overflow texts, third-party APIs similar to Java API classes could also have been mistakenly considered.

Threats to external validity relate to the generalizablity of a technique. So far, we experimented using API classes from only standard Java libraries.
However, since our technique mainly exploits co-occurrence between keywords and APIs, the technique can be easily adapted for API recommendation in other programming domains.


Threats to construct validity relate to suitability of evaluation metrics. We use Top-K Accuracy and Reciprocal Rank which are widely used for evaluating recommendation systems \cite{feature,pick}. 
The remaining two metrics are well known in information retrieval, and are also frequently used by studies \cite{feature,conngraph,portfolio} relevant to our work.
This  confirms no or little threats to construct validity.

\section{Related Work}\label{sec:related}\vspace{-.1cm}
\textbf{API Recommendation:} Existing studies on API recommendation accept one or more natural language queries, and recommend relevant API classes and methods by analyzing
code surfing behaviour of the developers and API invocation chains \cite{portfolio}, API dependency graphs \cite{conngraph}, feature request history or API documentations \cite{feature}, and library usage patterns \cite{librec}. 
\citet{portfolio} first propose \emph{Portfolio} that recommends relevant API methods for a code search query by employing natural language processing, indexing and graph-based algorithms (\eg\ PageRank).
\citet{conngraph} improve upon \emph{Portfolio}, and return a connected subgraph containing the most relevant APIs by employing further sophisticated graph-mining and textual similarity techniques.
\citet{feature} recommend relevant API methods to assist the implementation of an incoming feature request by analyzing request history and textual similarity between API details and the request text.
In short, each of these relevant techniques above considers lexical similarity between a query and the signature or documentation of the API for collecting candidate APIs, which might not be always effective given that query formulation could be highly subjective.
On the other hand, we exploit two co-occurrence heuristics that are derived from crowdsourced knowledge for collecting the candidate APIs, which are found to be relatively more effective.
Co-occurrence heuristics overcome the vocabulary mismatch problem \cite{qeffect}, and provide a generic, both language and project independent solution.
Besides, we exploit the expertise of a large crowd of technical users from Stack Overflow for API recommendation which none of the past studies did.
We compare with two two variants of the state-of-the-art technique--\citeauthor{feature}, and readers are referred to Section \ref{sec:compare} for detailed comparison.
Since \citeauthor{feature} outperform \citeauthor{conngraph} as reported \cite{feature}, we compared only with \citeauthor{feature} for our validation.

\textbf{API Usage Pattern Recommendation:} 
\citet{parseweb} propose \emph{ParseWeb} that takes in a \emph{source object type} and a \emph{destination object type}, and returns a sequence of method invocations 
that serve as a solution which yields the destination object from the source object. \citet{mapo} take a query that describes the method or class of an API, and recommends a frequent sequence of method invocations for the API by analyzing hundreds of open source projects. 
\citet{suade} recommend a set of API methods that are relevant to a target method by analyzing the structural dependencies between the two sets.
Each of these techniques is relevant to our work since they recommend API methods. However, they operate on structured queries rather than natural language queries, and thus comparing ours with them is not feasible. Of course, we introduced two heuristics and exploited crowd knowledge for API recommendation which were not considered by any of the existing techniques.  This makes our contribution significantly different from all of them.

\section{Conclusion \& Future Work}\label{sec:conclusion}\vspace{-.2cm}
To summarize, we propose a novel technique--RACK-- that translates a natural language code search query into a ranked list of relevant APIs. 
It exploits two novel heuristics derived from crowdsourced knowledge for collecting the relevant APIs.
Experiments using 150 code search queries from three Java tutorial sites show that RACK recommends APIs with about 79\% Top-10 accuracy which is highly promising.
Comparison with two variants of the state-of-the-art technique shows that our technique outperforms both of them in accuracy, precision and recall by a large margin.
While that technique is project-sensitive, ours is generic, project independent, and it exploits invaluable crowdsourced knowledge.
In future, we plan to apply the co-occurrence heuristics in recommending for other software maintenance tasks such as concept location and traceability link recovery.

\balance


\bibliographystyle{plainnat}
\setlength{\bibsep}{0pt plus 0.3ex}
\bibliography{sigproc}  

\begin{thebibliography}{34}
\providecommand{\natexlab}[1]{#1}
\providecommand{\url}[1]{\texttt{#1}}
\expandafter\ifx\csname urlstyle\endcsname\relax
  \providecommand{\doi}[1]{doi: #1}\else
  \providecommand{\doi}{doi: \begingroup \urlstyle{rm}\Url}\fi

\bibitem[cdf()]{cdf}
Theoretical {CDF}.
\newblock URL \url{http://stats.stackexchange.com/questions/132652}.

\bibitem[de()]{de}
{Stack {E}xchange {D}ata {E}xplorer}.
\newblock URL \url{http://data.stackexchange.com/stackoverflow}.

\bibitem[jav({\natexlab{a}})]{java2s}
Java2s: {J}ava {T}utorials, {\natexlab{a}}.
\newblock URL \url{http://java2s.com}.

\bibitem[jav({\natexlab{b}})]{javadb}
Java{DB}: {J}ava {C}ode {E}xamples, {\natexlab{b}}.
\newblock URL \url{http://www.javadb.com}.

\bibitem[jso()]{jsoup}
Jsoup: {J}ava {HTML} {P}arser.
\newblock URL \url{http://jsoup.org}.

\bibitem[kod()]{kodejava}
Kode{J}ava: {J}ava {E}xamples.
\newblock URL \url{http://kodejava.org}.

\bibitem[luc()]{lucene}
Apache {L}ucene {C}ore.
\newblock URL \url{https://lucene.apache.org/core}.

\bibitem[ref()]{reflection}
Reflections {L}ibrary.
\newblock URL \url{https://code.google.com/p/reflections}.

\bibitem[sta()]{stable}
Stable {J}ava version.
\newblock URL \url{http://stackoverflow.com/questions/6223493}.

\bibitem[sto()]{stop}
Stopword {L}ist.
\newblock URL \url{https://code.google.com/p/stop-words}.

\bibitem[Bacchelli et~al.(2010)Bacchelli, Lanza, and Robbes]{email}
A.~Bacchelli, M.~Lanza, and R.~Robbes.
\newblock Linking e-{M}ails and {S}ource {C}ode {A}rtifacts.
\newblock In \emph{Proc. ICSE}, pages 375--384, 2010.

\bibitem[Bajracharya and Lopes(2012)]{cslog}
S.~K. Bajracharya and C.~V. Lopes.
\newblock Analyzing and {M}ining a {C}ode {S}earch {E}ngine {U}sage {L}og.
\newblock \emph{Empirical Softw. Engg.}, 17\penalty0 (4-5):\penalty0 424--466,
  2012.

\bibitem[Brandt et~al.(2009)Brandt, Guo, Lewenstein, Dontcheva, and
  Klemmer]{twostudy}
J.~Brandt, P.J. Guo, J.~Lewenstein, M.~Dontcheva, and S.R. Klemmer.
\newblock Two {S}tudies of {O}pportunistic {P}rogramming: {I}nterleaving {W}eb
  {F}oraging, {L}earning, and {W}riting {C}ode.
\newblock In \emph{Proc. SIGCHI}, pages 1589--1598, 2009.

\bibitem[Chan et~al.(2012)Chan, Cheng, and Lo]{conngraph}
W.~Chan, H.~Cheng, and D.~Lo.
\newblock Searching {C}onnected {API} {S}ubgraph via {T}ext {P}hrases.
\newblock In \emph{Proc. FSE}, pages 10:1--10:11, 2012.

\bibitem[Dagenais and Robillard(2012)]{robi}
B.~Dagenais and M.P. Robillard.
\newblock Recovering {T}raceability {L}inks between an {API} and its {L}earning
  {R}esources.
\newblock In \emph{Proc. ICSE}, pages 47--57, 2012.

\bibitem[Gosling et~al.(2012)Gosling, Joy, Steele, and Bracha]{gosling}
J.~Gosling, B.~Joy, G.~Steele, and G.~Bracha.
\newblock The {J}ava {L}anguage {S}pecification: {J}ava {SE} 7 {E}dition.
\newblock 2012.

\bibitem[Haiduc and Marcus(2011)]{qeffect}
S.~Haiduc and A.~Marcus.
\newblock On the {E}ffect of the {Q}uery in {IR}-based {C}oncept {L}ocation.
\newblock In \emph{Proc. ICPC}, pages 234--237, June 2011.

\bibitem[Harris(1968)]{cooccur}
Z.~Harris.
\newblock Mathematical {S}tructures in {L}anguage {C}ontents.
\newblock 1968.

\bibitem[Kevic and Fritz(2014{\natexlab{a}})]{kevic}
K.~Kevic and T.~Fritz.
\newblock Automatic {S}earch {T}erm {I}dentification for {C}hange {T}asks.
\newblock In \emph{Proc. ICSE}, pages 468--471, 2014{\natexlab{a}}.

\bibitem[Kevic and Fritz(2014{\natexlab{b}})]{kevicdict}
K.~Kevic and T.~Fritz.
\newblock A {D}ictionary to {T}ranslate {C}hange {T}asks to {S}ource {C}ode.
\newblock In \emph{Proc. MSR}, pages 320--323, 2014{\natexlab{b}}.

\bibitem[McMillan et~al.(2011)McMillan, Grechanik, Poshyvanyk, Xie, and
  Fu]{portfolio}
C.~McMillan, M.~Grechanik, D.~Poshyvanyk, Q.~Xie, and C.~Fu.
\newblock Portfolio: {F}inding {R}elevant {F}unctions and their {U}sage.
\newblock In \emph{Proc. ICSE}, pages 111--120, 2011.

\bibitem[Mihalcea and Tarau(2004)]{rada}
R.~Mihalcea and P.~Tarau.
\newblock Textrank: {B}ringing {O}rder into {T}exts.
\newblock In \emph{Proc. EMNLP}, pages 404--411, 2004.

\bibitem[Ponzanelli et~al.(2014)Ponzanelli, Bavota, Di~Penta, Oliveto, and
  Lanza]{prompter}
L.~Ponzanelli, G.~Bavota, M.~Di~Penta, R.~Oliveto, and M.~Lanza.
\newblock Mining {S}tack{O}verflow to {T}urn the {IDE} into a {S}elf-confident
  {P}rogramming {P}rompter.
\newblock In \emph{Proc. MSR}, pages 102--111, 2014.

\bibitem[Rahman et~al.(2014)Rahman, Yeasmin, and Roy]{surfclipse}
M.~M. Rahman, S.~Yeasmin, and C.~K. Roy.
\newblock Towards a {C}ontext-{A}ware {IDE}-{B}ased {M}eta {S}earch {E}ngine
  for {R}ecommendation about {P}rogramming {E}rrors and {E}xceptions.
\newblock In \emph{Proc. CSMR-WCRE}, pages 194--203, 2014.

\bibitem[Rigby and Robillard(2013)]{peter}
P.C. Rigby and M.P. Robillard.
\newblock Discovering {E}ssential {C}ode {E}lements in {I}nformal
  {D}ocumentation.
\newblock In \emph{Proc. ICSE}, pages 832--841, 2013.

\bibitem[Thongtanunam et~al.(2015)Thongtanunam, Kula, Yoshida, Iida, and
  Matsumoto]{pick}
P.~Thongtanunam, R.~G. Kula, N.~Yoshida, H.~Iida, and K.~Matsumoto.
\newblock Who {S}hould {R}eview my {C}ode ?
\newblock In \emph{Proc. SANER}, pages 141--150, 2015.

\bibitem[Thummalapenta and Xie(2007)]{parseweb}
S.~Thummalapenta and T.~Xie.
\newblock Parseweb: {A} {P}rogrammer {A}ssistant for {R}eusing {O}pen {S}ource
  {C}ode on the {W}eb.
\newblock In \emph{Proc. ASE}, pages 204--213, 2007.

\bibitem[Thung et~al.(2013{\natexlab{a}})Thung, Lo, and Lawall]{librec}
F.~Thung, D.~Lo, and J.~Lawall.
\newblock Automated {L}ibrary {R}ecommendation.
\newblock In \emph{Proc. WCRE}, pages 182--191, 2013{\natexlab{a}}.

\bibitem[Thung et~al.(2013{\natexlab{b}})Thung, Wang, Lo, and Lawall]{feature}
F.~Thung, S.~Wang, D.~Lo, and J.~Lawall.
\newblock Automatic {R}ecommendation of {API} {M}ethods from {F}eature
  {R}equests.
\newblock In \emph{Proc. ASE}, pages 290--300, 2013{\natexlab{b}}.

\bibitem[Vassallo et~al.(2014)Vassallo, Panichella, Di~Penta, and
  Canfora]{codes}
C.~Vassallo, S.~Panichella, M.~Di~Penta, and G.~Canfora.
\newblock Codes: {M}ining {S}ource {C}ode {D}escriptions from {D}evelopers
  {D}iscussions.
\newblock In \emph{Proc. ICPC}, pages 106--109, 2014.

\bibitem[Warr and Robillard(2007)]{suade}
F.~W. Warr and M.~P. Robillard.
\newblock Suade: {T}opology-{B}ased {S}earches for {S}oftware {I}nvestigation.
\newblock In \emph{Proc. ICSE}, pages 780--783, 2007.

\bibitem[Wong et~al.(2013)Wong, Yang, and Tan]{autocomment}
E.~Wong, J.~Yang, and L.~Tan.
\newblock {AutoComment: Mining Question and Answer sites for Automatic Comment
  Generation}.
\newblock In \emph{Proc. ASE}, pages 562--567, 2013.

\bibitem[Xie and Pei(2006)]{mapo}
T.~Xie and J.~Pei.
\newblock {MAPO}: {M}ining {A}pi {U}sages from {O}pen {S}ource {R}epositories.
\newblock In \emph{Proc. MSR}, pages 54--57, 2006.

\bibitem[Yuan et~al.(2014)Yuan, Lo, and Lawall]{wordsim}
T.~Yuan, D.~Lo, and J.~Lawall.
\newblock Automated {C}onstruction of a {S}oftware-specific {W}ord {S}imilarity
  {D}atabase.
\newblock In \emph{Proc. CSMR-WCRE}, pages 44--53, 2014.

\end{thebibliography}
%
%
\end{document}